\documentclass[A4paper,twocolumn]{article}
\usepackage{amssymb}

\usepackage{graphicx}
\usepackage{amsmath}

\begin{document}

\title{Realignment Entanglement Criterion for Continuous Bipartite Symmetric
Quantum States }
\author{Xiao-yu Chen$^1$, Li-zhen Jiang$^1$, Ping Yu$^2$, Mingzhen Tian$^3$ \\
%EndAName
{\small {$^1$ College of Information and Electronic Engineering, Zhejiang
Gongshang University, Hangzhou, Zhejiang 310018, China }}\\
{\small {$^2$ Department of Physics and Astronomy, University of Missouri,
Columbia, Missouri 65211, USA }}\\
{\small {$^3$ School of Physics, Astronomy and Computational Sciences,
George Mason University, Fairfax, Virginia 22030, USA}}}
\date{}
\maketitle

\begin{abstract}
The realignment entanglement criterion of bipartite non-Gaussian states is
obtained by applying the technique of functional analysis. The realignment
criterion is given as one inequality in contrast to the infinitive number of
inequalities based on the moments. We give the necessary and sufficient
condition of inseparability for non-Gaussian states prepared by photon
subtraction or addition from symmetric Gaussian states. The entanglement
criterion of non-Gaussian states evolved in thermal noise and amplitude
damping environment is also obtained.

PACS number(s): 03.67.Mn; 03.65.Ud\newline
\end{abstract}

\textit{Introduction}---The inseparabilty of a non-Gaussian state is a
longer standing problem in continuous variable (CV) quantum information
theory. For the entanglement of a quantum state, various criteria had been
developed, among them are Peres-Horodecki positive partial transpose (PPT)
criterion \cite{Peres}, the uncertainty criterion \cite{Hofmann} \cite{Ghune}%
, the entropy criterion \cite{Walborn} and the cross norm and realignment
criterion \cite{Rodolph} \cite{KChen}. The entanglement criteria initial
developed for the case of discrete variable systems are also applicable to
continuous variable systems. The entanglement criteria had been obtained for
Gaussian states \cite{Duan} \cite{Simon} \cite{Wang} \cite{Giedke} which are
completely characterized by the first and second moments. For $1\times 1$%
(each partite with one mode) \cite{Duan} \cite{Simon}and $1\times N$ \cite
{Werner} Gaussian states, the entanglement condition is necessary and
sufficient. It ceases to be sufficient for all higher mode systems \cite
{Werner}. The inseparability of non-Gaussian states are \cite{SV} \cite
{Hillery} \cite{Agarwal} \cite{Chen07} \cite{Nha} under intensify researches
after the success for Gaussian states. A hierarchy of entanglement
conditions were derived using high order moments \cite{SV}. Although strict
in the sense of PPT, they are not convenient for use except limited to low
orders of moments. For each particular CV state, a large number of moments
should be calculated when the orders of moments get higher and higher. The
same problem exists for Fock space PPT criterion \cite{Chen07}. The newly
founded entanglement criterion based on teleportation fidelity relies on the
choice of test function \cite{Nha}. Until now, to our best knowledge, all
the entanglement conditions obtained for non-Gaussian states are sufficient
but not necessary .

We will use realignment criterion to study the entanglement of non-Gaussian
states. The method is to write a non-Gaussian state as the functional
derivative of a Gaussian state (Gaussian kernel). We first treat the
realignment operation to the Gaussian kernel in Fock space, the operation
then is expressed as a transform of the second moments. With the help of the
generating functional and its characteristic function, the trace norm of the
realigned non-Gaussian density matrix then is simplified. If the trace norm
is greater than $1$, the state is entangled. For some non-Gaussian states
with experiment interests, the realignment entanglement criterion is
necessary and sufficient. A non-Gaussian pure state will evolve to a
non-Gaussian mixed state when it interacts with environment. There are
debates \cite{Allegra} on the conclusions based on numerical calculations of
the critical separable time of the evolved non-Gaussian states. We derive
the strict formula of the realignment criterion to clarify the problem.

\textit{Realignment entanglement criterion of Gaussian States}---In Fock
basis, a two mode Gaussian state with nullified first moments is written as
\cite{Chen07}
\begin{eqnarray*}
\rho _{k_1k_2,m_1m_2} &=&\frac 1{\sqrt{k_1!k_2!m_1!m_2!\det (\gamma ^{\prime
})}}\frac{\partial ^{k_1}}{\partial s_1^{k_1}}\frac{\partial ^{k_2}}{%
\partial s_2^{k_2}}\frac{\partial ^{m_1}}{\partial s_1^{\prime m_1}}\frac{%
\partial ^{m_2}}{\partial s_2^{\prime m_2}} \\
&&\times \exp \left. \left\{ \frac 12(s,s^{\prime })[\sigma _1\otimes
I_2+\beta ](s,s^{\prime })^T\right\} \right| _{s=s^{\prime }=0},
\end{eqnarray*}
where $s=(s_1,s_2),s^{\prime }=(s_1^{\prime },s_2^{\prime });$ $\sigma _i$
are Pauli matrices, and $I_2$ is the $2\times 2$ identity matrix. Here $%
\gamma ^{\prime }=\gamma +\sigma _1\otimes \frac{I_2}2,$where $\gamma $ is
the complex covariance matrix (CCM) of the state $\rho $ defined by $\gamma
_{ij}=\frac 12Tr[\rho (\Delta \mathcal{C}_i\Delta \mathcal{C}_j+\Delta
\mathcal{C}_j\Delta \mathcal{C}_i)],$ with $\mathcal{C}=(a_1^{\dagger
},a_2^{\dagger },a_1,a_2)$, $\Delta \mathcal{C=C-}Tr(\rho \mathcal{C})$ and $%
a_i,a_i^{\dagger }$are the annihilation and creation operators of the two
modes.The matrix $\beta $ is
\begin{equation}
\beta =(\sigma _3\otimes I_2)\gamma ^{\prime }{}^{-1}(\sigma _3\otimes I_2).
\label{wee0}
\end{equation}
The first moments of the state can be removed by local operations, so $%
Tr(\rho \mathcal{C})=0$ is assumed.

The realignment operation transforms the state $\rho $ to $\rho ^R$ ($\rho ^R
$ is usually not a quantum state) such that \cite{Lupo}
\begin{equation}
\rho _{k_1k_2,m_1m_2}^R=\rho _{k_1m_1,k_2m_2}.  \label{wee1}
\end{equation}
The realignment operation leads to a chain of interchanges or substitutions.
It can be expressed as the interchange of the subscripts: $%
k_2\Leftrightarrow m_1$, then can be expressed as the interchange of the
parameters: $s_2\Leftrightarrow s_1^{\prime },$ the operation can be further
expressed with a change of the matrix $\beta $ to $\beta _R$, we have
\begin{equation}
\beta _R=Z\beta Z+I_2\otimes \sigma _1-\sigma _1\otimes I_2  \label{wee2}
\end{equation}
where $Z$ is a $4\times 4$ matrix with entries $Z_{11}=Z_{23}=Z_{32}=Z_{44}=1
$ and all the other entries are zeros. Denote
\begin{equation}
\gamma _R^{\prime -1}=Z^{\prime }\gamma ^{\prime -1}Z^{\prime }+I_2\otimes
\sigma _1+\sigma _1\otimes I_2,  \label{wee3}
\end{equation}
with $Z^{\prime }=(\sigma _3\otimes I_2)Z(\sigma _3\otimes I_2).$ \ Using
local operations, the covariance matrix of a two mode Gaussian state can be
brought to its standard form \cite{Duan}. We assume $\gamma $ to be a
standard CCM. A symmetric two mode Gaussian state is defined by $Tr(\rho
a_1^{\dagger }a_1)=Tr(\rho a_2^{\dagger }a_2).$ Then for a symmetric
Gaussian state
\begin{equation}
\gamma =\left[
\begin{array}{llll}
0 & c_1 & b_0 & c_2 \\
c_1 & 0 & c_2 & b_0 \\
b_0 & c_2 & 0 & c_1 \\
c_2 & b_0 & c_1 & 0
\end{array}
\right] ,  \label{wee3a}
\end{equation}
where the real parameters $c_1=Tr(\rho a_1^{\dagger }a_2^{\dagger })=Tr(\rho
a_1a_2),$ $c_2=Tr(\rho a_1^{\dagger }a_2)=Tr(\rho a_2^{\dagger }a_1),$ $%
b_0=Tr(\rho a_1^{\dagger }a_1)+\frac 12.$ To simplify notations, we denote
matrix $\gamma $ by its first row as $\gamma =\{0,c_1,b_0,c_2\}.$ Denote $%
b=b_0+\frac 12,$then $\gamma ^{\prime }=\{0,c_1,b,c_2\}.$ Hence $\gamma
^{\prime -1}=K_1I_2\otimes I_2+K_2I_2\otimes \sigma _1+K_3\sigma _1\otimes
I_2+K_4\sigma _1\otimes \sigma _1$ can be denoted as
\begin{equation}
\gamma ^{\prime -1}=\{K_1,K_2,K_3,K_4\},  \label{wee4}
\end{equation}
with $K_1=\frac{2bc_1c_2}{\Delta ^2},$ $K_2=\frac{c_1}{\Delta ^2}%
(c_1^2-c_2^2-b^2),$ $K_3=\frac b{\Delta ^2}(b^2-c_1^2-c_2^2),$ $K_4=\frac{c_2%
}{\Delta ^2}(c_2^2-c_1^2-b^2),$ where $\Delta ^2=\det (\gamma ^{\prime }).$
We have $Z^{\prime }\gamma ^{\prime -1}Z^{\prime }=\{K_1,-K_3,-K_2,K_4\},$
hence
\begin{equation}
\gamma _R^{\prime -1}=\{K_1,1-K_3,1-K_2,K_4\}.  \label{wee5}
\end{equation}
Notice that $\rho ^R$ may not be a quantum state, however it is proportional
to a quantum Gaussian state $\rho ^{\prime }.$ There are two cases. The
first case is that $\gamma _R$ is a valid CCM, then $\rho ^{\prime }$ is a
quantum Gaussian state with $\gamma _R$ being its CCM, and $\rho ^R=\sqrt{%
\frac{\det (\gamma _R^{\prime })}{\det (\gamma ^{\prime })}}\rho ^{\prime },$%
so we get
\begin{equation}
Tr\rho ^R=\sqrt{\frac{\det (\gamma _R^{\prime })}{\det (\gamma ^{\prime })}}=%
\sqrt{\frac{\det (\gamma ^{\prime -1})}{\det (\gamma _R^{\prime -1})}}.
\label{wee6}
\end{equation}
A direct calculation shows that $\det (\gamma _R^{\prime -1})=\det (\gamma
^{\prime -1})+$ $\Omega ,$ where $\Omega
=4[(K_2-K_3)^2-(K_1-K_4)^2](1-K_2-K_3).$ Further we have $\Omega =\frac
4{\Delta ^2}[(b+c_1)^2-(b+c_1)-c_2^2].$ Thus the realignment entanglement
criterion $Tr\sqrt{\rho ^R\rho ^{R\dagger }}=Tr\rho ^R>1$ is equivalent to $%
(b+c_1)^2-(b+c_1)-c_2^2<0.$ That is
\begin{equation}
(b_0+c_1+c_2)(b_0+c_1-c_2)<\frac 14.  \label{wee7}
\end{equation}
The second case is that $\gamma _R$ is not a valid CCM, then we try the
operator $\rho _\Pi ^R=\rho ^R\Pi $ which may have all nonnegative singular
values such that $Tr\sqrt{\rho ^R\rho ^{R\dagger }}=Tr\rho _\Pi ^R$, where
\[
\Pi =\sum_{n_1,n_2}(-1)^{n_1+n_2}\left| n_1n_2\right\rangle \left\langle
n_1n_2\right| .
\]
Then $\rho _\Pi ^R=\sqrt{\frac{\det (\gamma _{R1}^{\prime })}{\det (\gamma
^{\prime })}}\rho _\Pi ^{\prime },$ where the CCM of the quantum state $\rho
_\Pi ^{\prime }$ is $\gamma _{R\Pi }$ with $\gamma _{R\Pi }^{\prime
-1}=Z^{\prime \prime }\gamma ^{\prime -1}Z^{\prime \prime }+I_2\otimes
\sigma _1+\sigma _1\otimes I_2,$with $Z^{\prime \prime }=(\sigma _3\otimes
I_2)Z^{\prime }.$ Then
\[
\gamma _{R\Pi }^{\prime -1}=\{K_1,1-K_3,1+K_2,-K_4\}.
\]
The effect of appending $\Pi $ corresponds to reversing both signs of $c_1$%
and $c_2$. The realignment entanglement criterion is
\begin{equation}
(b_0-c_1-c_2)(b_0-c_1+c_2)<\frac 14.  \label{wee8}
\end{equation}
The condition for the entanglement of a symmetric Gaussian state is that
either (\ref{wee7}) or (\ref{wee8}) should be fulfilled. The result of
realignment entanglement criterion is exactly the same obtained previously
for a symmetric Gaussian state by other criteria \cite{Duan} \cite{Simon}
\cite{SV}.

The original realignment criterion for entanglement is $Tr\sqrt{%
\rho ^{R\dagger }\rho ^R}>1.$ In general case, we may not have $Tr\sqrt{\rho
^{R\dagger }\rho ^R}=Tr\rho ^R.$ Note that $\rho ^R$ has the singular value
decomposition $\rho ^R=U\Lambda V^{\dagger }$, with unitary operators $U$, $V
$ and diagonal operator $\Lambda =diag\{\Lambda _1,\Lambda _2,\ldots \}.$ So
$Tr\sqrt{\rho ^{R\dagger }\rho ^R}=Tr\left| \Lambda \right| .$ Let $%
U^{\prime }=V^{\dagger }U$ be another unitary operator, we have $Tr\rho
^R=Tr(U^{\prime }\Lambda )=\sum_jU_{jj}^{\prime }\Lambda _j\leq \sum_j\left|
U_{jj}^{\prime }\right| \left| \Lambda _j\right| \leq \sum_j\left| \Lambda
_j\right| $, the last inequality is due to the fact that $\left|
U_{jj}^{\prime }\right| \leq 1$ for any unitary operator $U^{\prime }$.
Hence if $Tr\rho ^R$ is real, we have $Tr\rho ^R\leq Tr\sqrt{\rho ^{R\dagger
}\rho ^R}.$ So if $Tr\rho ^R>1,$ then $Tr\sqrt{\rho ^{R\dagger }\rho ^R}\geq
Tr\rho ^R>1.$

For a non-symmetric state with standard covariance matrix, the inseparable
criterion can also be written in the form of (\ref{wee7}) or (\ref{wee8})
with $b_0$ replaced by $\sqrt{b_1b_2},$ and $b_i=Tr(\rho a_i^{\dagger
}a_i)+\frac 12$ for $i=1,2.$

\textit{Realignment} \textit{inseparability of Non-Gaussian State}---For the
inseparability of non-Gaussian states, we introduce the method of generating
functional. Let
\begin{equation}
W=e^{\varepsilon a^{\dagger }}e^{\xi a}\rho e^{\eta a^{\dagger }}e^{\zeta a},
\label{wee9}
\end{equation}
where $\varepsilon ,\xi ,\eta ,\zeta $ are two dimensional real parameters,
we denote $\xi _1a_1+\xi _2a_2$ as $\xi a\ $and so on, $\rho $ is the
initial two mode Gaussian state with CCM $\gamma $. Then non-Gaussian state $%
\rho _N$ can be obtained from functional $W$ by derivations. $\rho _N=%
\mathcal{O}W$, where $\mathcal{O}$ is the operation of derivations on the
parameters $\varepsilon ,\xi ,\eta ,\zeta $ and later setting all these
parameters to be zeros. In Fock basis,
\begin{eqnarray*}
W_{k_1k_2,m_1m_2} &=&\frac{e^{\eta \xi ^T+\frac 12(\xi ,\eta )\beta (\xi
,\eta )^T}}{\sqrt{k_1!k_2!m_1!m_2!\det (\gamma ^{\prime })}}\frac{\partial
^{k_1}}{\partial s_1^{k_1}}\frac{\partial ^{k_2}}{\partial s_2^{k_2}}\frac{%
\partial ^{m_1}}{\partial s_1^{\prime m_1}} \\
&&\frac{\partial ^{m_2}}{\partial s_2^{\prime m_2}}\exp \left[ \frac
12(s,s^{\prime })[\sigma _1\otimes I_2+\beta ](s,s^{\prime })^T\right] \\
&&\left. \cdot \exp \left\{ (s,s^{\prime })[\beta (\xi ,\eta )^T+(\eta
+\varepsilon ,\xi +\zeta )^T\right\} \right| _{s=s^{\prime }=0},
\end{eqnarray*}
The realignment is the interchange of $k_2\Leftrightarrow m_1$ such that the
realigned functional operator $W^R$ has entries $%
W_{k_1k_2,m_1m_2}^R=W_{k_1m_1,k_2m_2}.$ The operator $W^R$ can be put in the
form of
\[
W^R=e^{\varepsilon ^Ra^{\dagger }}e^{\xi ^Ra}\rho ^Re^{\eta ^Ra^{\dagger
}}e^{\zeta ^Ra},
\]
the new parameters are $(\xi ^R,\eta ^R)=(\xi ,\eta )Z;$ $(\varepsilon
^R,\zeta ^R)=(\varepsilon ,\zeta )Z.$ Denote the realignment of the
non-Gaussian state as $\rho _N^R=\mathcal{O}W^R$. The realignment
entanglement criterion is $Tr\rho _N^R=\mathcal{O}TrW^R>1.$ The
characteristic function of the operator $W$ is $\chi ^W(\mu )=Tr[W$ $%
\mathcal{D}(\mu )],$ where $\mathcal{D}(\mu )=\exp (\mu a^{\dagger }-\mu
^{*}a)$ is the displacement operator, with $\mu $ being a two dimensional
complex parameter. After a lengthy calculation, we obtain
\begin{eqnarray}
\chi ^W(\mu ) &=&P\exp [-\frac 12(\mu ,\mu ^{*})\gamma (\mu ,\mu
^{*})^T]\exp \{(-\mu ,\mu ^{*})  \nonumber \\
&&\cdot [\beta ^{-1}(\varepsilon +\eta ,\xi +\zeta )^T+(\xi ,\eta )^T]\}.
\label{wee11}
\end{eqnarray}
where
\[
P=\exp [-\frac 12(\varepsilon +\eta ,\xi +\zeta )\beta ^{-1}(\varepsilon
+\eta ,\xi +\zeta )^T-\zeta \eta ^T-\varepsilon \xi ^T-\eta \xi ^T].
\]
Similarly, we can obtain the characteristic function $\chi ^{WR}(\mu )$ of
the operator $W^R,$ then $TrW^R=\chi ^{WR}(0)=\sqrt{\frac{\det (\gamma
_R^{\prime })}{\det (\gamma ^{\prime })}}P^R,$ with
\begin{eqnarray*}
P^R &=&\exp [-\frac 12(\varepsilon ^R+\eta ^R,\xi ^R+\zeta ^R)\beta
_R^{-1}(\varepsilon ^R+\eta ^R,\xi ^R+\zeta ^R)^T \\
&&-\zeta ^R\eta ^{RT}-\varepsilon ^R\xi ^{RT}-\eta ^R\xi ^{RT}].
\end{eqnarray*}
The realignment inseparable criterion for a non-Gaussian state is
\begin{equation}
\sqrt{\frac{\det (\gamma _R^{\prime })}{\det (\gamma ^{\prime })}}\mathcal{O}%
P^R>1.  \label{att1}
\end{equation}

The first example is the non-Gaussian state prepared by photon subtracting
from or adding to a symmetric Gaussian state. For a photon subtracted state $%
\rho _{-}=c_{-}a_1a_2\rho a_1^{\dagger }a_2^{\dagger }$ or a photon added
state $\rho _{+}=c_{+}a_1^{\dagger }a_2^{\dagger }\rho a_1a_2,$, where $%
c_{_{\mp }}=[(b_0\mp \frac 12)^2+c_1^2+c_2^2]^{-1}$ are the normalizations.
A simple technique to calculate $\gamma _R^{\prime }$ is $\gamma _R^{\prime
}=H(H\gamma _R^{\prime -1}H)^{-1}H,$ where $H=I_2\otimes H_2,$ with $H_2$
being $2\times 2$ Hadamard matrix. Thus
\begin{eqnarray}
\gamma _R^{\prime } &=&\frac 12\{c_2(\tau -1),(b_0+c_1)\tau -(b_0-c_1),
\nonumber \\
&&(b_0+c_1)\tau +(b_0-c_1)+1,c_2(\tau +1)\},  \label{att2}
\end{eqnarray}
where $\tau =\frac 1{4[(b_0+c_1)^2-c_2^2]}.$ We have
\begin{equation}
\mathcal{O}P^R=\frac{c_{_{\mp }}}2[(b_0-c_1\mp \frac 12)^2+((b_0+c_1)\tau
\mp \frac 12)^2+\frac{c_2^2}2(\tau +1)^2],  \label{att3}
\end{equation}
for photon subtraction and addition. If the Gaussian kernel is on the
separable boundary, namely, $\tau =1,$ then $\mathcal{O}P^R=1.$ This
indicates that $\rho _{\mp }$ are separable. In fact, if the Gaussian kernel
$\rho $ is separable thus it is probability mixture of product states, the
definitions of $\rho _{\pm \text{ }}$ show that $\rho _{\pm \text{ }}$ are
also probability mixtures of product states. We confirm that if the Gaussian
kernel is separable, $\rho _{\mp }$ are separable too. If the Gaussian
kernel state $\rho $ is entangled, namely, $\sqrt{\frac{\det (\gamma
_R^{\prime })}{\det (\gamma ^{\prime })}}=\sqrt{\tau }>1,$ we can show $%
\mathcal{O}P^R>1$. Notice that$\left. \mathcal{O}P^R\right| _{\tau =1}=1,$
we need to show $\left. \mathcal{O}P^R\right| _{\tau >1}>\left. \mathcal{O}%
P^R\right| _{\tau =1},$ this is true from (\ref{att3}). The necessary and
sufficient entanglement condition of $\rho _{\mp }$is exactly the same as
that of its Gaussian Kernel state. Consider states prepared by subtracting
or adding $m$ photons from each mode of the standard symmetric Gaussian
state $\rho $. The states prepared are $\rho _{-m}=$ $c_{_{-}m}a_1^ma_2^m%
\rho a_1^{\dagger m}a_2^{\dagger m}$ and $\rho _{+m}=$ $c_{_{-}m}a_1^{%
\dagger m}a_2^{\dagger m}\rho a_1^ma_2^m.$ The states are separable when
their Gaussian kernels are separable by the same reasoning for $\rho _{\pm
\text{ }}$. Assume a function
\[
f(m,V)=\left. \frac{\partial ^4}{\partial \kappa _1^m\partial \kappa
_2^m\partial \kappa _3^m\partial \kappa _2^m}\right| _{\kappa =0}\exp
[-\frac 12\kappa V\kappa ^T]
\]
for a matrix $V=\{V_1,V_2,V_3,V_4\},$ where $\kappa =(\kappa _1,\ldots
,\kappa _4).$ Notice that $Tr\rho _N=Tr\mathcal{O}W$ $=\mathcal{O}\chi (0)=%
\mathcal{O}P=1$, we have
\begin{equation}
\mathcal{O}P^R=\frac{f(m,\gamma _R\mp \frac 12\sigma _1\otimes I_2)}{%
f(m,\gamma \mp \frac 12\sigma _1\otimes I_2)}.  \label{att4}
\end{equation}
for states $\rho _{\mp m}.$ For the states $\rho _{\mp 2},$ we have
\begin{eqnarray*}
f(2,V) &=&(V_1^2+2(V_2^2+V_3^2+V_4^2))^2+32V_1V_2V_3V_4 \\
&&+8(V_2^2V_3^2+V_2^2V_4^2+V_3^2V_4^2).
\end{eqnarray*}
Numerical results show that $\mathcal{O}P^R>1$ for $\tau >1.$ Hence the
necessary and sufficient condition of inseparability for states $\rho _{\mp
2}$ is (\ref{wee7}) too.

The second example is a non-Gaussian state $\rho _N=p\rho _1+(1-p)\rho _2,$
the probability mixture of an entangled Gaussian state $\rho _1$ with a
separable Gaussian state $\rho _2.$ For simplicity, consider the case of
symmetric Gaussian state with $c_2=0$, namely the two mode squeezed thermal
state (TMST). Denote $w=b_0+c_1-\frac 12$, then the entanglement criterion
is simply $w<0$. Thus $w_1\in (-\frac 12,0)$ for $\rho _1$ and $w_2\geq 0$
for $\rho _2.$ The inseparable criteria of the second moments \cite{Simon} ,
the Fock space \cite{Chen07} and the realignment for $\rho _N$ are
\begin{eqnarray*}
pw_1+(1-p)w_2 &<&0, \\
pw_1+(1-p)w_2+w_1w_2 &<&0, \\
pw_1+(1-p)w_2+2w_1w_2 &<&0,
\end{eqnarray*}
respectively. Hence realignment criterion is the best among the three.

\textit{Inseparability of time evolution states}---For the evolved state, we
will use subscript $t$ in the following to specify its operator and
parameters, while the initial state is not with such a subscript. In an
amplitude damping and thermal noise channel, the time evolution of the
characteristic function for any quantum state is \cite{Chen06}
\begin{equation}
\chi _t(\mu )=\chi (\mu e^{-\frac{\Gamma t}2})\exp [-(\widetilde{n}+\frac
12)(1-e^{-\Gamma t})\left| \mu \right| ^2],  \label{wee12}
\end{equation}
where $\Gamma $ is the amplitude damping coefficient and $\widetilde{n}$ is
the thermal noise, we assume the two mode undergo the same environment.
Notice that $\chi ^W(\mu )$ in (\ref{wee11}) is proportional to the
characteristic function of a two mode Gaussian state with first moments. The
time evolution of $\chi ^W(\mu )$ then is
\begin{eqnarray}
\chi _t^W(\mu ) &=&P\exp [-\frac 12(\mu ,\mu ^{*})\gamma _t(\mu ,\mu
^{*})^T]\exp \{(-\mu ,\mu ^{*})  \nonumber \\
&&\cdot [\beta _t^{-1}(\varepsilon _t+\eta _t,\xi _t+\zeta _t)^T+(\xi
_t,\eta _t)^T]\}  \label{wee13}
\end{eqnarray}
Where
\begin{eqnarray*}
\gamma _t &=&e^{-\Gamma t}\gamma +(\widetilde{n}+\frac 12)(1-e^{-\Gamma
t})\sigma _1\otimes I_2; \\
\beta _t &=&(\sigma _3\otimes I_2)(\gamma _t+\sigma _1\otimes \frac{I_2}%
2){}^{-1}(\sigma _3\otimes I_2).
\end{eqnarray*}
The equation $e^{-\frac 12\Gamma t}[\beta ^{-1}(\varepsilon +\eta ,\xi
+\zeta )^T+(\xi ,\eta )^T]=$ $\beta _t^{-1}(\varepsilon _t+\eta _t,\xi
_t+\zeta _t)^T+(\xi _t,\eta _t)^T$ leads to the parameter transformation
\begin{eqnarray*}
(\varepsilon _t,\zeta _t)^T &=&e^{-\frac 12\Gamma t}\beta _t\beta
^{-1}(\varepsilon ,\zeta )^T; \\
(\eta _t,\xi _t)^T &=&e^{-\frac 12\Gamma t}(\beta _t^{-1}+\sigma _1\otimes
I_2)^{-1}(\beta ^{-1}+\sigma _1\otimes I_2)(\eta ,\xi )^T.
\end{eqnarray*}
Denote
\begin{eqnarray*}
P_t &=&\exp [-\frac 12(\varepsilon _t+\eta _t,\xi _t+\zeta _t)\beta
_t^{-1}(\varepsilon _t+\eta _t,\xi _t+\zeta _t)^T \\
&&-\zeta _t\eta _t^T-\varepsilon _t\xi _t^T-\eta _t\xi _t^T],
\end{eqnarray*}
then the time evolved $W$ operator is
\begin{equation}
W_t=PP_t^{-1}e^{\varepsilon _ta^{\dagger }}e^{\xi _ta}\rho _te^{\eta
_ta^{\dagger }}e^{\zeta _ta}.  \label{wee14}
\end{equation}
where $\rho _t$ is a Gaussian state with CCM $\gamma _t$. The realignment
operation on time dependent functional $W_t$ will lead to
\begin{eqnarray}
W_t^R &=&PP_t^{-1}e^{\varepsilon _t^Ra^{\dagger }}e^{\xi _t^Ra}\rho
_t^Re^{\eta _t^Ra^{\dagger }}e^{\zeta _t^Ra}  \nonumber \\
&=&PP_t^{-1}\sqrt{\frac{\det (\gamma _{Rt}^{\prime })}{\det (\gamma
_t^{\prime })}}e^{\varepsilon _t^Ra^{\dagger }}e^{\xi _t^Ra}\rho
_t^{^{\prime }}e^{\eta _t^Ra^{\dagger }}e^{\zeta _t^Ra}.  \label{wee15}
\end{eqnarray}
where
\begin{eqnarray*}
\gamma _t^{\prime } &=&\gamma _t+\sigma _1\otimes \frac{I_2}2; \\
\gamma _{Rt}^{\prime -1} &=&Z^{\prime }\gamma _t^{\prime -1}Z^{\prime
}+I_2\otimes \sigma _1+\sigma _1\otimes I_2; \\
(\xi _t^R,\eta _t^R) &=&(\xi _t,\eta _t)Z; \\
(\varepsilon _t^R,\zeta _t^R) &=&(\varepsilon _t,\zeta _t)Z.
\end{eqnarray*}
We have used the fact that $\rho _t^R$ is not a quantum state, however it is
proportional to a quantum state $\rho _t^{^{\prime }}$ with CCM $\gamma
_{Rt}.$ Here $\gamma _{Rt}$ is the realignment transformation of the time
dependent CCM $\gamma _t$. The characteristic function of $W_t^R$ can be
obtained accordingly following the process from Eq.(\ref{wee9}) to Eq.(\ref
{wee11}), however, as we will show below, what we need to know is $%
Tr(W_t^R)=\chi _t^{WR}(0)$, which is
\begin{equation}
\chi _t^{WR}(0)=PP_t^{-1}\sqrt{\frac{\det (\gamma _{Rt}^{\prime })}{\det
(\gamma _t^{\prime })}}P_t^R,  \label{wee16}
\end{equation}
where
\begin{eqnarray*}
P_t^R &=&\exp [-\frac 12(\varepsilon _t^R+\eta _t^R,\xi _t^R+\zeta
_t^R)\beta _{Rt}^{-1}(\varepsilon _t^R+\eta _t^R,\xi _t^R+\zeta _t^R)^T \\
&&-\zeta _t^R(\eta _t^R)^T-\varepsilon _t^R(\xi _t^R)^T-\eta _t^R(\xi
_t^R)^T]
\end{eqnarray*}
with $\beta _{Rt}=(\sigma _3\otimes I_2)(\gamma _{Rt}^{\prime
}){}^{-1}(\sigma _3\otimes I_2).$

The characteristic function of the initial state is

\[
\chi _N(\mu )=Tr[\mathcal{O}W\mathcal{D}(\mu )]=\mathcal{O}Tr[W\mathcal{D}%
(\mu )]=\mathcal{O}\chi ^W(\mu ).
\]
The second equality is due to the fact that the trace operation is on the
creation and annihilation operators $a^{\dagger }$and $a,$ and $\mathcal{O}$
is about the parameters $\varepsilon ,\xi ,\eta ,\zeta .$ The evolution of
the characteristic function is a transformation of $\mu ,$ which is
independent of the operations on parameters $\varepsilon ,\xi ,\eta ,\zeta ,$
hence
\[
\chi _{Nt}(\mu )=\mathcal{O}\chi _t^W(\mu ).
\]
We further have the evolved non-Gaussian state
\begin{eqnarray*}
\rho _{Nt} &=&\int \prod_i\left[ \frac{d^2\mu _i}\pi \right] \chi _{Nt}(\mu )%
\mathcal{D}(-\mu ) \\
&=&\mathcal{O}\int \prod_i\left[ \frac{d^2\mu _i}\pi \right] \chi _t^W(\mu )%
\mathcal{D}(-\mu )=\mathcal{O}W_t,
\end{eqnarray*}
and the realignment of the non-Gaussian state $\rho _{Nt}^R=\mathcal{O}%
W_t^R. $ If all of the singular values of $\rho _{Nt}^R$ are non-negative,
then $Tr\sqrt{\rho _{Nt}^R\rho _{Nt}^{R\dagger }}=Tr\rho _{Nt}^R=\mathcal{O}%
TrW_t^R=\mathcal{O}\chi _t^{WR}(0).$ The realignment entanglement criterion
for a bipartite non-Gaussian state is
\begin{equation}
\sqrt{\frac{\det (\gamma _{Rt}^{\prime })}{\det (\gamma _t^{\prime })}}%
\mathcal{O}PP_t^{-1}P_t^R>1  \label{wee17}
\end{equation}
If some of the singular values of $\rho _{Nt}^R$ are negative, then $Tr\sqrt{%
\rho _{Nt}^R\rho _{Nt}^{R\dagger }}>Tr\rho _{Nt}^R$. We still have (\ref
{wee17}) as the sufficient condition of entanglement. It can be further
written as
\begin{equation}
\sqrt{\frac{\det (\gamma _{Rt}^{\prime })}{\det (\gamma _t^{\prime })}}%
\mathcal{O}\exp [-\frac 12\mathbf{v}\left[
\begin{array}{ll}
M & M^{\prime } \\
M^{\prime } & M^{\prime }
\end{array}
\right] \mathbf{v}^T]>1.  \label{wee18}
\end{equation}
where $\mathbf{v=}(\varepsilon ,-\zeta ,\eta ,-\xi ),$
\begin{eqnarray*}
M &=&\gamma ^{\prime }+e^{-\Gamma t}\gamma ^{\prime }(-\gamma _t^{\prime
-1}+\gamma _t^{\prime -1}Z^{\prime }\gamma _{Rt}^{\prime }Z^{\prime }\gamma
_t^{\prime -1})\gamma ^{\prime }, \\
M^{\prime } &=&\gamma ^{\prime \prime }+e^{-\Gamma t}\gamma ^{\prime \prime
}(-\gamma _t^{\prime \prime -1}+\gamma _t^{\prime \prime -1}Z\gamma
_{Rt}^{\prime \prime }Z\gamma _t^{\prime \prime -1})\gamma ^{\prime \prime },
\end{eqnarray*}
with
\begin{eqnarray*}
\gamma ^{\prime \prime } &=&\gamma ^{\prime }-\sigma _1\otimes I_2, \\
\gamma _t^{\prime \prime } &=&\gamma _t^{\prime }-\sigma _1\otimes I_2, \\
\gamma _{Rt}^{\prime \prime } &=&\gamma _{Rt}^{\prime }-\sigma _1\otimes I_2.
\end{eqnarray*}

\textit{Inseparability of photon subtracted and added two mode squeezed
vacuum states---}The initial state of photon subtracted state is $\rho
_S=c_sa_1a_2\rho a_1^{\dagger }a_2^{\dagger },$ where $\rho =\left| \psi
\right\rangle \left\langle \psi \right| ,$
\[
\left| \psi \right\rangle =\sqrt{1-\lambda ^2}\sum_{n=0}^\infty \lambda
^n\left| nn\right\rangle
\]
is a two mode squeezed vacuum state with $\lambda =\tanh (r),$ where $r$ is
the squeeze parameter. The normalization constant is $c_s=\frac{(1-\lambda
^2)^2}{\lambda ^2(1+\lambda ^2)}.$ The state is anti-normally ordered with
operator
\[
\mathcal{O}=\left. c_s\frac{\partial ^4}{\partial \xi _1\partial \xi
_2\partial \eta _1\partial \eta _2}\right| _{\xi =\eta =0}.
\]
The initial state of photon added state is $\rho _A=c_aa_1^{\dagger
}a_2^{\dagger }\rho a_1a_2,$ with $c_a=\frac{(1-\lambda ^2)^2}{1+\lambda ^2}%
. $ The state is normally ordered with
\[
\mathcal{O}=\left. c_a\frac{\partial ^4}{\partial \varepsilon _1\partial
\varepsilon _2\partial \zeta _1\partial \zeta _2}\right| _{\varepsilon
=\zeta =0}.
\]
The CCM of the two mode squeezed vacuum state is $\gamma =\{0,-\frac \lambda
{1-\lambda ^2}$ $,\frac 1{1-\lambda ^2}-\frac 12,0\},$ so $\gamma ^{\prime
}=\{0,-\frac \lambda {1-\lambda ^2}$ $,\frac 1{1-\lambda ^2},0\}\ $and $%
\gamma _t^{\prime }=\{0,-\Lambda ,N,0\}$ where
\begin{eqnarray*}
\Lambda &=&\frac \lambda {1-\lambda ^2}e^{-\Gamma t}, \\
N &=&\frac{\lambda ^2}{1-\lambda ^2}e^{-\Gamma t}+\widetilde{n}(1-e^{-\Gamma
t})+1.
\end{eqnarray*}
We have
\[
\gamma _{Rt}^{\prime -1}=\{0,1-\frac N{N^2-\Lambda ^2},1-\frac \Lambda
{N^2-\Lambda ^2},0\},
\]
so
\begin{equation}
\sqrt{\frac{\det (\gamma _{Rt}^{\prime })}{\det (\gamma _t^{\prime })}}%
=\frac 1{2(N-\Lambda )-1}.  \label{wee19}
\end{equation}
The detail calculation shows
\begin{eqnarray*}
-\gamma _t^{\prime -1}+\gamma _t^{\prime -1}Z^{\prime }\gamma _{Rt}^{\prime
}Z^{\prime }\gamma _t^{\prime -1} &=&-\gamma _t^{\prime \prime -1}+\gamma
_t^{\prime \prime -1}Z\gamma _{Rt}^{\prime \prime }Z\gamma _t^{\prime \prime
-1} \\
&=&-\frac 1{2N-2\Lambda -1}\{0,1,1,0\}
\end{eqnarray*}
for the evolution of two mode squeezed vacuum state. For the photon added
state, we have $\eta =\xi =0,$ the matrix
\[
M=\frac 1{1+\lambda }\{0,-\frac \lambda {1-\lambda }-Q,\frac 1{1-\lambda
}-Q,0\},
\]
with $Q=\frac 1{(1+\lambda )[2(N-\Lambda )-1]}e^{-\Gamma t}.$ The
inseparable criterion is reduced to
\begin{equation}
\frac 1{2(N-\Lambda )-1}\frac{c_a}{(1+\lambda )^2}[(\frac \lambda {1-\lambda
}+Q)^2+(\frac 1{1-\lambda }-Q)^2]>1  \label{wee20}
\end{equation}
For the photon subtract state we have $\varepsilon =\zeta =0$, the matrix $%
M^{\prime }=\frac \lambda {1+\lambda }\{0,-\frac 1{1-\lambda }-\lambda Q,$ $%
\lambda (\frac 1{1-\lambda }-Q),0\}.$ The inseparable criterion is
\begin{equation}
\frac 1{2(N-\Lambda )-1}\frac{c_s\lambda ^2}{(1+\lambda )^2}[(\frac
1{1-\lambda }+\lambda Q)^2+\lambda ^2(\frac 1{1-\lambda }-Q)^2]>1
\label{wee21}
\end{equation}
The entanglement conditions based on second moments \cite{Simon} \cite{SV}
are
\begin{equation}
N-\Lambda +\frac{(1-\lambda )^3}{1-\lambda ^4}e^{-\Gamma t}<1  \label{wee22}
\end{equation}
for photon added state and
\begin{equation}
N-\Lambda -\frac{\lambda (1-\lambda )^3}{1-\lambda ^4}e^{-\Gamma t}<1
\label{wee23}
\end{equation}
for photon subtracted state. For photon subtraction, the critical time
separability is comparable for the two criteria. Realignment criterion is
better than the second moment criterion at the higher initial squeezing
parameter side ($\lambda \gtrsim 0.3$). For photon addition, realignment
criterion is better for almost all the parameters.

\textit{Summary}---We derive a simple formula of the realignment criterion
for bipartite continuous variable bipartite system. The main contribution of
the paper is to propose generating functional and its characteristic
function in calculating the trace of the realigned density operator for
arbitrary two-mode non-Gaussian state such that the realignment criterion is
applicable to continuous variable system. We reduce the trace to the product
of two factors. One factor is determined by the second moments of the
Gaussian kernel of the non-Gaussian state. The other is a derivative of an
exponential function. For symmetric Gaussian states, the realignment
criterion gives the same condition as the other criteria. For photon added
states and subtracted states prepared from symmetric Gaussian states, the
separable criterion is necessary and sufficient. It is just the separable
criterion of the source symmetric Gaussian states. The condition can be
extended to bi-photon (possibly to multiple photon) subtraction or addition
case. The realignment criterion is better than second moment criterion and
one photon Fock space criterion for detecting the entanglement of the
mixture of entangled TMST and separable TMST. The complete formula of the
trace is derived for any two mode continuous variable system evolved in
thermal noise and amplitude damping environment.

XYC and LZJ thank the support of the National Natural Science Foundation of
China (Grant No. 60972071).

\end{document}